\documentclass[francais]{gretsi}

\usepackage[utf8]{inputenc}

\usepackage[T1]{fontenc}
\usepackage[french]{babel}

\usepackage{amsmath, amssymb, amsfonts}

\usepackage{mdframed}
\usepackage{acronym}
\usepackage{graphicx}
\usepackage{epstopdf}
\usepackage{amsthm}
\usepackage{amsfonts}
\usepackage{dsfont,bbm}
\usepackage{url}
\usepackage{color}

\newcommand{\diag}{\mathrm{diag}}

\usepackage{fancyhdr}
 
\graphicspath{{graphics/}}

\acrodef{iid}[i.i.d.]{independent identically distributed}
\acrodef{wrt}[w.r.t.]{with respect to}

\acrodef{SINR}[SINR]{signal-to-interference plus noise ratio}
\acrodef{SNR}[SNR]{signal-to-noise ratio}
\acrodef{DMC}[DMC]{discrete memoryless channel}
\acrodef{BSC}[BSC]{binary symmetric channel}
\acrodef{KKT}[KKT]{Karush-Kuhn-Tucker}
\acrodef{NPC}[NPC]{Nash-equilibrium power control}
\acrodef{SPC}[SPC]{Semi-coordinated power control}
\acrodef{CCPC}[CCPC]{Costless-communication power control}
\acrodef{CPC}[CPC]{Coded power control}

\newtheorem{theorem}{Theorem}[section]

\newtheorem{proposition}[theorem]{Proposition}

\begin{document}


\titre{Improving MIMO channel estimation via receive power feedback}

\auteurs{
  \auteur{Chao}{Zhang}{}{1}
  \auteur{Hang}{Zou}{}{2}
  \auteur{Samson}{Lasaulce}{}{3}
  \auteur{Lucas}{Saludjian}{}{4}
}

\affils{
  \affil{}{$^1$Central South University \& $^2$TII \& $^3$CRAN-Nancy \& $^4$RTE France
  }
}


\resume{Estimer l'\'{e}tat du canal est un probl\`{e}me important dans les r\'{e}seaux sans fil. \`{A} cette fin, il importe d'exploiter toutes les informations disponibles pour am\'{e}liorer autant que possible la pr\'{e}cision de l'estimation de canal. Il s'av\`{e}re que le probl\`{e}me de l'exploitation de l'information associ\'{e}e au feedback de puissance re\c{c}u (par exemple, l'indicateur de force du signal re\c{c}u -RSSI-) n'a pas \'{e}t\'{e} identifi\'{e} et r\'{e}solu; dans cette configuration, on suppose que l'\'{e}metteur re\c{c}oit des commentaires de tous les r\'{e}cepteurs pr\'{e}sents. Pour r\'{e}soudre ce probl\`{e}me, des outils d'estimation classiques peuvent \^{e}tre utilis\'{e}s. L'utilisation du MMSE correspondant est toujours b\'{e}n\'{e}fique, tandis que la pertinence de l'utilisation de l'estimateur MAP d\'{e}pendrait du rapport signal sur bruit (SNR) de fonctionnement.}

\abstract{Estimating the channel state is known to be an important problem in wireless networks. To this end, it matters to exploit all the available information to improve channel estimation accuracy as much as possible. It turns out that the problem of exploiting the information associated with the receive power feedback (e.g., the received signal strength indicator -RSSI-) has not been identified and solved; in this setup, the transmitter is assumed to receive feedback from all the receivers in presence. As shown in this paper, to solve this problem, classical estimation tools can be used. Using the corresponding MMSE is shown to be always beneficial, whereas the relevance of using the MAP estimator would depend on the operating SNR. }

\maketitle


\section{Introduction}
\label{sec:intro}

The acquisition of channel information is essential to optimize the system performance in many wireless networks such as orthogonal frequency-division multiplexing (OFDM) and multiple input multiple output (MIMO) (see e.g., \cite{Li-2002,Fang-TWC-2017,Choi-TSP-2014,Inter-TWC-2019}). Generally the channel information is not perfectly known in practical systems and needs to be estimated by sending pilot sequences \cite{caire-tit-2010} or using blind channel estimation \cite{blind-2002}. Improving the quality of channel estimation is a well studied problem for both academic researchers as well as engineers in the communication industry.  {Recent studies of channel estimation through downlink training concentrated on massive MIMO systems.  The overhead of training sequence is huge in such systems with a large number of antennas which makes the channel estimation even more challenging.   The behavior of MMSE estimator
of  low-rank channel covariance matrix is studied in \cite{Fang-TWC-2017}.  It is shown that training overhead can be substantially reduced by exploiting the low-rank property of the channel covariance matrix in the asymptotic low-noise regime. In \cite{Choi-TSP-2014}, an open and closed-loop training
framework using successive channel prediction/estimation at the user for FDD massive MIMO systems is proposed. Similarly, a small amount of
feedback is required in low signal-to-noise ratio (SNR) regime with the immense transmit antennas or with inaccurate prior channel estimates setup for closed-loop systems. More interestingly, in \cite{Inter-TWC-2019}, downlink
beamforming training, although increases pilot overhead and
introduces additional pilot contamination, improves significantly
the achievable downlink rate.} Most of the works aims at finding efficient training schemes to improve the channel estimate, especially in low SNR regime. However, apart from the training phase, the channel estimation can also be enhanced during data transmission, which is the purpose of this paper. 

In this paper, by exploiting the receive power feedback \cite{Zhang-TWC-2017}, also referred to as received signal strength indicator (RSSI), we propose a novel technique to better estimate the channel in MIMO systems. {Since feedback resources are extremely limited in practical systems, it is important to efficiently use these power domain feedback resources. For example, authors in \cite{Panda-ICACCI-2016} consider a sub-optimal scheduling scheme for a multiuser MIMO (MU-MIMO) to decrease the load on the feedback. Received signal-to-noise-plus-interference-ratio (SINR) at each user is quantized for both fixed thresholds and adaptable ones both achieving a remarkable reduction of feedback bits. It is worth mentioning that the usage of such information could be highly case-depending.  Authors in \cite{Su-TVT-2015} show that the quantization method for a single cell could highly degrade the performance of coordinated multipoint (CoMP) system for quantization error vector being no longer isotropic.} To this end, we propose to use receive power feedback at a much lower rate compared to the classical channel state information (CSI)  feedback rate. We will focus on two  specific estimators: minimum mean square error (MMSE) and maximum a posteriori (MAP) estimators. Both estimators have been widely used to acquire CSI but only through pilot training, e.g., in \cite{caire-tit-2010}\cite{vincent-book}.

\section{Problem statement}

We consider a MIMO broadcast channel consisting of a base station (BS) equipped with $N$  transmit antennas and $K$ user equipments (UEs) with single antenna. The channel model can be described by 
\begin{equation}
y_j = \mathbf{h}_{j}^{\mathrm{H}} \mathbf{x} + z_j
\end{equation}
where $y_j$ represents the received signal at UE $j$,  $\mathbf{h}_j=(h_{1j},\dots,h_{Nj})^{\mathrm{T}}\in \mathbb{C}^N$ is the vector of channel coefficients from BS to UE $j$, $\mathbf{x}=(x_1,\dots,x_N)^{\mathrm{T}}$ is the vector of independent transmitted symbols at BS and $z_j$ is the received Gaussian noise at UE $j$ following $z_j\sim \mathcal{CN}(0,N_0)$. The channel matrix, $\mathbf{H}=[\mathbf{h}_1,\dots,\mathbf{h}_K]$, is assumed to follow a block fading model, that is, $\mathbf{H}$ remains constant by block and more precisely over each transmitted frame. Indeed, like modern cellular networks standards, we assume that the transmission take places in the form of frames, each frame being itself divided into $M$ time-slots, and each time slot comprises $T$ symbols. In frequency division duplex (FDD) wireless systems, one can assume that  each UE $j\in\{1,\dots,K\}$ estimates $\mathbf{h}_{j}$ from downlink training symbols and feeds the estimated channel information back to the BS. This is the setting we assume here. The conventional approach to estimate the channel is to use pilot signals and training in the signal domain only. Assume orthogonal pilot signals with power $P$. If $\beta$ symbols are used for training, the observation at Receiver $j$ that is, $\bold{s}_j=(s_{1j},\dots,s_{Nj})^T$ is given by
\begin{equation}
\bold{s}_j = \sqrt{\beta P} \mathbf{h}_j  + \bold{z}_j
\end{equation}
where the estimation noise $\bold{z}_j=(z_{1j},z_{2j},\dots,z_{Nj})^T\sim \mathcal{CN}(0,N_0\bold{I}_N)$.  The signal-to-noise ratio (SNR) in the training phase is defined as $\mathrm{SNR}=\frac{\beta P}{N_0}$. The observations $\bold{s}_j$ can be used to estimate $\mathbf{h}_j$. The approach retained in this paper is to de-noise the above estimates not only by using the channel statistics (as done in previous works such as \cite{lasaulce-vtc-2001}) but also by using some information that can be acquired in the power domain, namely the receive power feedback samples sent by the UEs to the BS. Indeed, in wireless systems such as 3G and 5G systems, for each transmitted time slot $m'$, the BS receives a feedback sample of the power received by UE $j$. This feedback is used e.g., for power control or scheduling at the BS. Our key observation is that this information can also be used to improve channel estimation, and more precisely the modulus information accuracy. Suppose the transmitted symbols at BS are not correlated, i.e. $\mathbb{E}[x_i(m')x_j(m')^{\mathrm{H}}]=0$ for $i\neq j$, the average receive power at UE $j$ for time slot $m'$ can be written as
\begin{equation}
R_{j,m'}= \sum_{i=1}^N g_{ij} P_i(m') + N_0
\label{eq:RSSI_feedback}
\end{equation}
where $g_{ij}=|h_{ij}|^2$ represents the channel gain and $P_i(t) = \mathbb{E}[|x_i(m')|^2]$ represents the transmit power of the $i$-th antenna at time-slot $m'$. Our goal is to find new estimates $\bold{h}_j^{\text{X}:m}$ by exploiting this power domain information, where $\text{X}\in\{\text{MMSE}, \text{MAP}\}$ indicates the type of considered estimation criterion, and $m$ the number of time slots exploited. Without loss of generality, we assume that $\bold{h}_j\sim \mathcal{CN}(0,\bold{D}_j)$ where $\bold{D}_j=\diag\{\sigma_{1j}^2,\dots,\sigma_{Nj}^2\}$.  
\section{Novel estimates with receive power feedback}
In this section, we explain our technique for the single band scenario. The extension to the multi-band case is straightforward since the estimation over each band can be processed independently.  Apart from the observation $\mathbf{s}_j$ during the downlink training phase, we use receive power feedback introduced by (\ref{eq:RSSI_feedback}) as an additional information to improve the estimation quality.
\subsection{MMSE estimator}

Denote by $\mathbf{R}_j^{(m)} = (R_{j,1},\dots,R_{j,m})$ the vector of random RSSI observed at $j$ over $m$ time slots.  Correspondingly, $\mathbf{r}_j^{(m)} = (r_{j,1},\dots,r_{j,m})$ denotes a vector of RSSI realizations. The optimal MMSE estimator knowing $\mathbf{r}_j^{(m)}$ can be written as
\begin{equation}\bold{h}_j^{\mathrm{MMSE}:m}=\bold{A}_j^{\mathrm{OPT:m}}\bold{s}_j\end{equation}
where $\bold{A}_j^{\mathrm{OPT:m}}$ is obtained from:
\begin{equation}
 \bold{A}_j^{\mathrm{OPT:m}}\in\underset{\bold{A}_j}{\arg\min} \quad \mathbb{E}_{\bold{h}_j,\bold{z}_j|\mathbf{R}_j^{(m)}=\mathbf{r}_j^{(m)}}[|\bold{A}_j\bold{s}_j-\bold{h}_j|^2]
\end{equation}
The corresponding optimal solution $\bold{A}_j^{\mathrm{OPT:m}}$ expressed as:
\begin{equation}
\bold{A}_j^{\mathrm{OPT:m}}=\mathbb{E}_{\bold{h}_j,\bold{z}_j|\mathbf{R}_j^{(m)}=\mathbf{r}_j^{(m)}}[\bold{h}_j\bold{s}_j^{\mathrm{H}}]\mathbb{E}_{\bold{h}_j,\bold{z}_j|\mathbf{R}_j^{(m)}=\mathbf{r}_j^{(m)}}[\bold{s}_j\bold{s}_j^{\mathrm{H}}]^{-1}
\label{eq:optimalAwithm}
\end{equation}

With our model, this could be rewritten as:

\begin{equation}
\bold{A}_j^{\mathrm{OPT:m}} = { \sqrt{\beta P}\mathbb{E}_{\bold{h}_j|\mathbf{R}_j^{(m)}=\mathbf{r}_j^{(m)}}[\bold{h}_j\bold{h}_j^{\mathrm{H}}] }\bold{W}^{-1}
\label{eq:optimalAwithm2}
\end{equation}
where $W={\beta P}\mathbb{E}_{\bold{h}_j|\mathbf{R}_j^{(m)}=\mathbf{r}_j^{(m)}}[\bold{h}_j\bold{h}_j^{\mathrm{H}}]+N_0\bold{I}_K$.

The $\mathbb{E}_{\bold{h}_j|\mathbf{R}_j^{(m)}=\mathbf{r}_j^{(m)}}[\bold{h}_j\bold{h}_j^{\mathrm{H}}]$ term in (\ref{eq:optimalAwithm2}) is not always easy to express. Therefore, we also provide details on how such an estimator can be designed in practice. First, it can be verified that the non-diagonal elements of $\mathbb{E}_{\bold{h}_j|\mathbf{R}_j^{(m)}=\mathbf{r}_j^{(m)}}[\bold{h}_j\bold{h}_j^{\mathrm{H}}]$ equal to zero, i.e.
\begin{equation}
\mathbb{E}_{\bold{h}_j|\mathbf{R}_j^{(m)}=\mathbf{r}_j^{(m)}}[h_{ij}h_{kj}^{*}]=0\quad(k\neq i)
\label{eq:zero-element}
\end{equation}
From (\ref{eq:zero-element}), it can be seen that $\bold{A}_j^{\mathrm{OPT:m}} $ is a diagonal matrix. Thus the MMSE estimator expresses:
\begin{equation}
\label{eq:mmse_gen}
{h}_{ij}^{\mathrm{MMSE}:m}=\frac{ \sqrt{\beta P}\mathbb{E}_{\bold{h}_j|\mathbf{R}_j^{(m)}=\mathbf{r}_j^{(m)}}[|{h}_{ij}|^2] }{{\beta P}\mathbb{E}_{\bold{h}_j|\mathbf{R}_j^{(m)}=\mathbf{r}_j^{(m)}}[|{h}_{ij}|^2]+N_0} s_{ij}
\end{equation}
The classical MMSE estimator can be obtained as a special case 
\begin{equation}
{h}_{ij}^{\mathrm{MMSE}:0}=\frac{ \sqrt{\beta P}\sigma_{ij}^2 }{{\beta P}\sigma_{ij}^2+N_0} s_{ij}.
\label{eq:MMSE_conventional}
\end{equation} 
Knowing the optimal estimator, we can calculate the conditional MSE (conditioned to $\mathbf{R}_j^{(m)}=\mathbf{r}_j^{(m)}$) as a result of using the MMSE given by (\ref{eq:optimalAwithm2}). This conditional MSE is denoted by $\mathrm{D}_{j}(\mathbf{r}_j^{(m)})$ when $m$ RSSI observations are available, and can be evaluated as $
\mathrm{D}_{j}(\mathbf{r}_j^{(m)})= \mathbb{E}_{|\mathbf{R}_j^{(m)}=\mathbf{r}_j^{(m)}}[|\bold{A}_j\bold{s}_j-\bold{h}_j|^2]$. Finally, the real MSE of the estimator with $m$ RSSI measurements can be expressed as $\Delta_{j;m}=\mathbb{E}_{\mathbf{r}_j^{(m)}}[\mathrm{D}_{j}(\mathbf{r}_j^{(m)})]$. We use $ D^*_{j} $ and $\Delta_{j;m}^*$ to denote the conditional and real MSEs when using the optimal estimator given in (\ref{eq:optimalAwithm2}). This yields:
\begin{align} 
 \label{eq:distor2}
\mathrm{D}_{j}^{*}(\mathbf{r}_j^{(m)})=&\sum_{i=1}^{N}\frac{N_0\mathbb{E}_{|\mathbf{R}_j^{(m)}=\mathbf{r}_j^{(m)}}[|h_{ij}|^2]}{\beta P\mathbb{E}_{|\mathbf{R}_j^{(m)}=\mathbf{r}_j^{(m)}}[|h_{ij}|^2]+N_0}
\\
 \label{eq:distor}
\Delta_{j;m}^{*}=&\mathbb{E}^{*}_{\mathbf{r}_j^{(m)}}[\mathrm{D}^{*}_{j}(\mathbf{r}_j^{(m)})]
\end{align}

Next, we prove that the performance in terms of the real MSE given in (\ref{eq:distor}) is at least as good as the classical estimator for the proposed estimator. We also prove that lesser MSE can be achieved if we use more feedback samples. We formalize this result with the following proposition.

\begin{proposition}
The MSE resulting from the proposed estimator in (\ref{eq:optimalAwithm2}) is a decreasing function of the number of RSSI feedbacks available at any receiver $j$, i.e., $\Delta^*_{j;m} \geq \Delta^*_{j;m+1}$. Additionally, the MSE is lower bounded by a constant
\begin{equation}
\Delta^*_{j;m} \geq \sum_{i=1}^N \mathbb{E}_{h_{ij}}\left[\frac{N_0|h_{ij}|^2}{\beta P|h_{ij}|^2+N_0}\right]
\end{equation}
\end{proposition}

\subsection{MAP estimator}

As with the MMSE estimator, we can use the information provided by RSSI measurements to propose a novel MAP estimator which reduces the MSE. Denoting $m$ RSSI measurements available at the $j$-th receiver by $\mathbf{r}_j^{(m)}$, the MAP estimate can be given by
{\footnotesize \begin{equation}
\begin{split}
\widehat{h}_{ij}^{\text{MAP:m}}&\in \arg \max_{h_{ij}} f_{h|s,r}(h_{ij}|{s}_{ij} ,\mathbf{r}_j^{(m)} )\\
&\in \arg \max_{h_{ij}} f_{s|h,r}(s_{ij}|h_{ij},\mathbf{r}_j^{(m)} )f_{r|h}(\mathbf{r}_j^{(m)}|h_{ij} )f_{h}(\mathbf{r}_j^{(m)}|h_{ij} )\\
&\in \arg \max_{h_{ij}} f_{r|h}(\mathbf{r}_j^{(m)}|h_{ij} )\exp(\frac{|s_{ij}-\sqrt{\beta P h_{ij}}|^2}{N_0})\exp(\frac{|h_{ij}|^2}{\sigma_{ij}^2})
\end{split}
\label{eq:newbasicmap}
\end{equation}}
When no RSSI measurements are available i.e., $m=0$, this reduces to the classical MAP as follows 
\begin{equation}
\widehat{h}_{ij}^{\text{MAP}:0} = \frac{\sqrt{\beta P}\sigma_{ij}^2}{N_0+ \beta P\sigma_{ij}^2 } s_{ij},
\label{eq:MAP_conventional}
\end{equation}
which also coincides with the classical MMSE. However, when the RSSI estimates are known, $f_{r|h}(\mathbf{r}_j^{(m)}|h_{ij} )$ becomes a Dirac function. Therfore, the optimization must be performed under the constraints that $R_{j}^{(1)}=r_{j}^{(1)},\dots,R_{j}^{(m)}=r_{j}^{(m)}$. The optimization problem can be rewritten as:

\begin{equation}
\begin{split}
&\underset{\bold{h}_j}{\min}\quad \frac{(\bold{s}_j-\sqrt{\beta P}\bold{h}_j)^H(\bold{s}_j-\sqrt{\beta P}\bold{h}_j)}{N_0}+{\bold{h}_j^H\bold{D}_j^{-1}\bold{h}_j}\\
&\mathrm{s.t.} \quad \bold{h}_j^{\mathrm{H}}\bold{P}(n)\bold{h}_j+N_0=r_{j}^{(n)}, \,\, \forall \,\, n \in \{1,2,\dots,m\}
\end{split}
\label{eq:MAP_op_origin}
\end{equation}
where  $\bold{P}(n)=\diag (P_1(n),\dots,P_N(n))$. This problem can be solved by using the Lagrange multiplier method. The optimal solution $\widehat{\bold{h}}^{\text{MAP:m}}_{ij}$ can be written as
\begin{equation}\label{eq:21}
\widehat{\bold{h}}^{\text{MAP:m}}_{ij}=\frac{\sqrt{\beta P}}{N_0(\frac{1}{\sigma^2_{ij}}+\frac{\beta P}{N_0}+\sum_{n=1}^{m}\lambda_n P_i(n))}{s}_{ij}
\end{equation}
where $\lambda_n$ can be obtained from the constraint $
\lambda_n ({\bold{h}_j}^{H}\bold{P}(n) {\bold{h}_j}-r_{j}^{(n)}+N_0)=0$ for all $n \in \{1,2,\dots,m\}$.
\textcolor{black}{If the number of timeslots $m$ is less than the number of antennas, then there are possibly infinite solutions for the above system of equations. To circumvent this problem,  we assume that we can perfectly know $|h_{ij}|^2$ or $g_{ij}$ for $i \in S$ by exploiting $m$ RSSI measurements where $S$ is defined as follows:} $S=\{i\in\{1,\dots,N\} |\quad |h_{ij}|^2\,\,\text{can be obtained from RSSI} \}$. Based on this relaxation, the RSSI measurements are treated as the measurements of channel gains/magnitude $ |h_{ij}|^2$, and the optimization problem defined by (\ref{eq:MAP_op_origin}) can be simplified as:
\begin{equation}
\begin{split}
&\underset{\bold{h}_j}{\min}\quad \frac{(\bold{s}_j-\sqrt{\beta P}\bold{h}_j)^H(\bold{s}_j-\sqrt{\beta P}\bold{h}_j)}{N_0}+{\bold{h}_j^H\bold{D}_j^{-1}\bold{h}_j}\\
&\mathrm{s.t.} \quad |h_{ij}|^2=g_{ij}, \,\,\forall \,\,i \in S
\end{split}
\end{equation}
The solution of the simplified problem can be found by solving the following equations:
\begin{equation}\label{eq:25}
\widehat{{h}}^{\text{MAP:m}}_{ij}=\frac{\sqrt{\beta P}}{N_0(\frac{1}{\sigma^2_{ij}}+\frac{\beta P}{N_0}+q_i)}{s}_{ij}\\
\end{equation}
\begin{equation}\label{eq:26}
\frac{\beta P |s_{ij}|^2 }{[(\frac{N_0}{\sigma_{ij}^2}+{\beta P})+N_0\lambda_i]^2}=g_{ij} , \,\,\forall \,\,i \in S
\end{equation}
where  $q_i=\lambda_i$ when $i \in S$ and $q_i=0$ otherwise.
With (\ref{eq:26}), for $i\in S$, we have:
\begin{equation}
\lambda_i=\frac{\displaystyle{\pm\sqrt{\frac{\beta P |s_{ij}|^2}{g_{ij}}}-(\frac{N_0}{\sigma_{ij}^2}+{\beta P})}}{N_0}.
\end{equation}
Therefore, the MAP estimator can be simplified and rewritten as:
\begin{equation}
\widehat{h}_{ij}^{\text{MAP:m}}=
\left\{
\begin{array}{ll}
\sqrt{g_{ij}} \frac{s_{ij}}{|s_{ij}|} & \hbox{ $i\in S$} \\\\
\widehat{h}_{ij}^{\text{MAP}:0} & \hbox{ otherwise} \\\\
\end{array}
\right.
\end{equation}
The proposed MAP will ensure the maximum a posteriori probability. However the minimization of the MSE is not always guaranteed. Indeed, it can be seen from the following proposition that the proposed MAP brings more MSE in the low SNR regime.
\begin{proposition}
Assuming all the channel gains can be obtained from the RSSI measurements, i.e., $S=\{1,\dots,N\}$, the following equalities can be derived
{\footnotesize \begin{equation}
\underset{\mathrm{SNR}\rightarrow 0}{\lim}\mathbb{E}_{h_{ij},z_{ij}}[|\widehat{h}_{ij}^{\text{MAP:m}}-h_{ij}|^2]=2\underset{\mathrm{SNR}\rightarrow 0}{\lim}\mathbb{E}_{h_{ij},z_{ij}}[|\widehat{h}_{ij}^{\text{MAP}}-h_{ij}|^2]
\end{equation}
\begin{equation}
\underset{\mathrm{SNR}\rightarrow \infty}{\lim}\mathbb{E}_{h_{ij},z_{ij}}[|\widehat{h}_{ij}^{\text{MAP:m}}-h_{ij}|^2]=\frac{1}{2}\underset{\mathrm{SNR}\rightarrow \infty}{\lim}\mathbb{E}_{h_{ij},z_{ij}}[|\widehat{h}_{ij}^{\text{MAP}}-h_{ij}|^2]
\end{equation}}
\end{proposition}


%
\section{Numerical results}
\label{sec:num}
In the simulations, we fix number of transmit antennas as$N=4$, number of users as $K=4$ and observe the distortion performance at different SNR. For ease of exposition, we assume $\bold{D}_j$ is an identity matrix in the simulations. i.e. $\bold{D}_j=\bold{I}_N$. As seen from (\ref{eq:MMSE_conventional}) and (\ref{eq:MAP_conventional}), the classical MMSE coincides with the classical MAP in our model. We define the relative MSE reduction as \begin{equation}
\frac{\Delta_{j;0}^{*}-\Delta_{j;m}^{\mathrm{X}*}}{\Delta_{j;0}^{*}}\times 100\%
\end{equation}
 with $\text{X}\in\{\text{MMSE}, \text{MAP}\}$. We assume that from each feedback, we can perfectly reconstruct one channel gain, i.e., with $m$ feedbacks, Receiver $j$ can acquire $g_{1j},\dots,g_{mj}$.  
 
 First, we study the influence of the prior information on the MMSE estimator. We compare the reduced MSE in terms of number of feedback samples for different SNR. Fig.~1 shows that the MSE can be mitigated by using the RSSI measurements and the MSE decreases with more measurements available. In the low SNR regime, the MMSE estimator does not perform well since $\mathbf{s}_j$ is multiplied by a constant matrix $\bold{A}_j$. Hence, even if we can improve the selection of $\bold{A}_j$ by knowing the RSSI, the MSE can not be reduced much due to the high noise level. On the other hand, in the high SNR regime, the classical MMSE estimator is already accurate. This results in limited gains for the high SNR regime too. Our technique can bring more improvements in the mid-SNR range. 
 
 We conduct a similar analysis for the new MAP estimator. The results are shown in Fig.~2. From Fig.~2, we see that unlike in the MMSE case, the MSE does not necessarily reduce with higher number of RSSI feedback samples. Indeed, as proved in Prop.~III.2, the MSE for the new MAP might actually increase in the low SNR regime. It has been checked that the degradation in estimation is close to 100\% when SNR is less than -40 dB. However, when SNR increases, the RSSI measurements can lead to a significant reduction in MSE and it can even outperform the new MMSE. 

 \begin{figure}[!h]
   \begin{center}
        \includegraphics[width=0.44\textwidth]{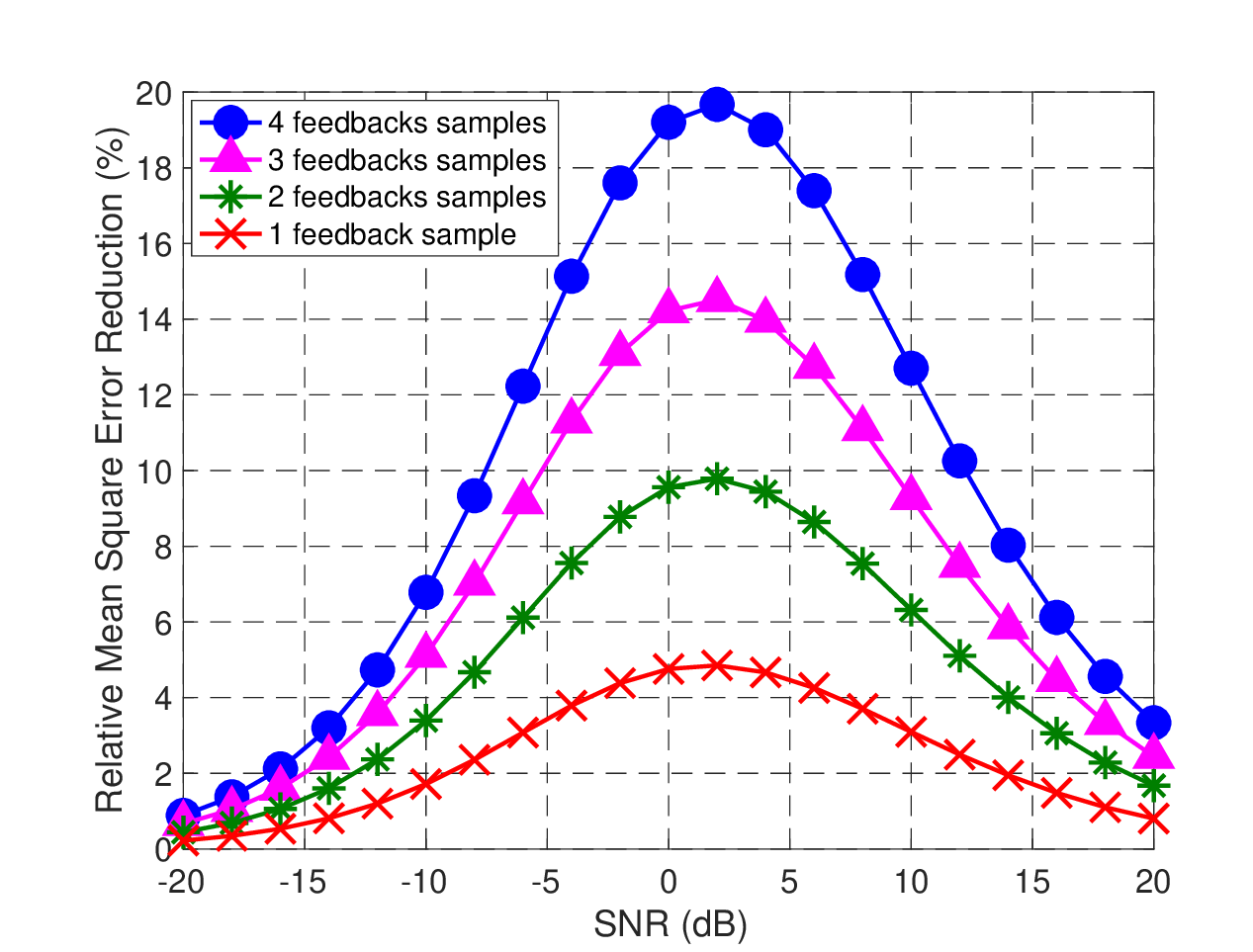}
    \end{center}
  \caption{The MMSE estimator brings a significant improvements in the range [-10dB,+10dB]}
   \label{fig:MMSE}
\end{figure}
\begin{figure}[!h]
   \begin{center}
        \includegraphics[width=0.44\textwidth]{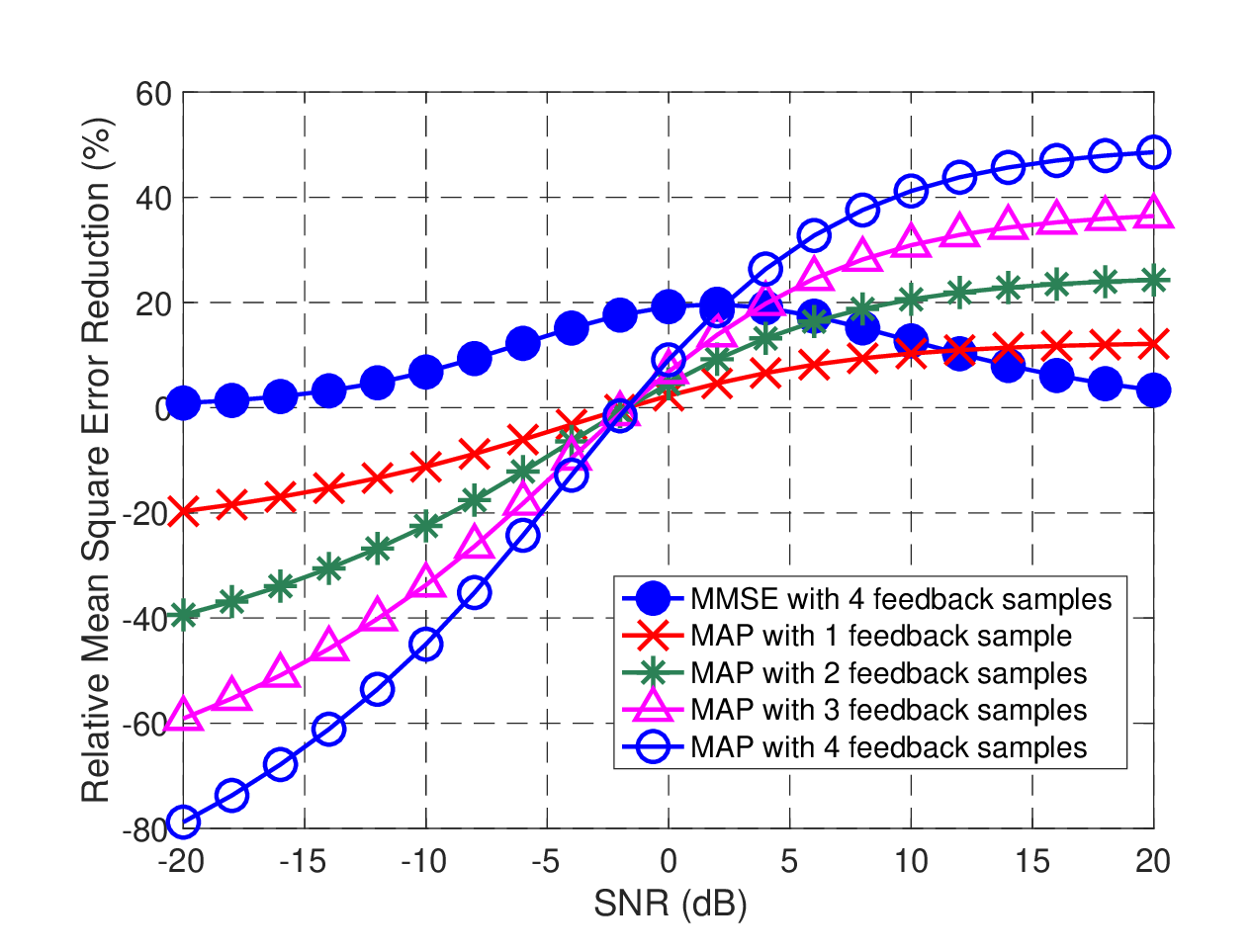}
    \end{center}
  \caption{The MAP estimator performs very well for sufficiently high SNRs}
   \label{fig:MAP}
\end{figure}

\section{Conclusion and future works}

Under some classical assumptions on the residual noise produced by least-squares estimates, it is known that the de-noising matrices obtained by using MMSE and MAP de-noising coincide \cite{lasaulce-vtc-2001}. In the presence of additional prior coming from the power domain, this result is no longer true, which explains that MMSE and MAP performance do not perform similarly here. In terms of MSE performance, it is recommended to use the proposed MMSE in the moderate SNR regime and the proposed MAP in the high SNR regime. By doing so, RSSI feedback can be used not only for resource allocation purposes (as currently done) but also for improving channel estimation. In future works, we intend to extend this framework to goal-oriented communication systems, where estimation quality is not only to recover the signal itself but also to serve the system to better accomplish the goal \cite{Zhang-AE-2021}\cite{Zhang-JSAC-2022}\cite{Zhang-IoTM-2022}.

\section*{Acknowledgement}
This work is partly supported by PGMO project fund from the Fondation Hadamard des Math\'ematiques.

\bibliographystyle{unsrt}


\end{document}